\documentclass{jkas0_2019}
\pdfoutput=1
\usepackage{graphicx}
\usepackage{amsmath}
\usepackage{supertabular}
\usepackage{multicol}
\usepackage{booktabs}
\usepackage{longtable}
\usepackage{gensymb}
\def\beginpage{11} % first page of article
\def\received{November 30, 2018} % date paper was received by JKAS
\def\accepted{January 10, 2019} % date of acceptance
\date{Received \received ; accepted \accepted}

\def\farcm{\hbox{$.\mkern-4mu^\prime$}}
\def\farcs{\hbox{$.\!\!^{\prime\prime}$}}
\title{
Intensive Monitoring Survey of Nearby Galaxies (IMSNG)
}

\author[1,2]{Myungshin~Im}
\author[1,2]{Changsu~Choi}
\author[1,2]{Sungyong Hwang}
\author[1,2]{Gu~Lim}
\author[1,2]{Joonho Kim}
\author[1,2]{Sophia Kim}
\author[1,2]{Gregory S. H. Paek}
\author[1,2]{Sang-Yun Lee}
\author[2]{Sung-Chul Yoon}
\author[3]{Hyunjin Jung}
\author[4]{Hyun-Il Sung}
\author[4]{Yeong-beom Jeon}
\author[5]{Shuhrat Ehgamberdiev}
\author[5]{Otabek Burhonov}
\author[5]{Davron Milzaqulov}
\author[5]{Omon Parmonov}
\author[2,6]{Sang Gak Lee}
\author[6]{Wonseok Kang}
\author[6,7]{Taewoo Kim}
\author[6]{Sun-gill Kwon}
\author[1,8]{Soojong Pak}
\author[1,8]{Tae-Geun Ji}
\author[1,8]{Hye-In Lee}
\author[1,8]{Woojin Park}
\author[9]{Hojae Ahn}
\author[9]{Seoyeon Byeon}
\author[9]{Jimin Han}
\author[10]{Coyne Gibson}
\author[10,11]{J. Craig Wheeler}
\author[10]{John Kuehne}
\author[12]{Chris Johns-Krull}
\author[13]{Jennifer Marshall}
\author[1,2]{Minhee Hyun}
\author[1,2]{Seong-Kook J. Lee}
\author[1,2]{Yongjung Kim}
\author[1,2]{Yongmin Yoon}
\author[1,2]{Insu Paek}
\author[1,2]{Suhyun Shin}
\author[1,2]{Yoon Chan Taak}
\author[2]{Juhyung Kang}
\author[14]{Seoyeon Choi}
\author[2]{Mankeun Jeong}
\author[2]{Moo-Keon Jung}
\author[15]{Hwara Kim}
\author[9]{Jisu Kim}
\author[16]{Dayae Lee}
\author[1]{Bomi Park}
\author[17]{Keunwoo Park}
\author[2]{Seong A O}
\affil[1]{Center for the Exploration of the Origin of the Universe, Department of Physics and Astronomy, Seoul National University, Gwanak-gu, Seoul 08826, Korea; \email{mim@astro.snu.ac.kr}}
\affil[2]{Astronomy Program, Department of Physics and Astronomy, Seoul National University, Gwanak-gu, Seoul 151-742, Korea}
\affil[3]{Department of Physics, Pohang University of Science and Technology, 77 Cheongam-Ro, Nam-Gu, Pohang, Gyeongbuk 37673, Korea}
\affil[4]{Korea Astronomy and Space Science Institute, 776 Daedeokdae-ro, Yuseong-gu, Daejeon 34055, Korea}
\affil[5]{Ulugh Beg Astronomical Institute, Uzbek Academy of Sciences, 33 Astronomical Street, Tashkent 700052, Uzbekistan}
\affil[6]{National Youth Space Center, Goheung, Jeollanam-do, 59567, Korea}
\affil[7]{Department of Astronomy and Space Science, Chungbuk National University, Cheongju-City, 28644, Korea}
\affil[8]{School of Space Research, Kyung Hee University, 1732 Deogyeong-daero, Giheung-gu, Yongin-si,  Gyeonggi-do 17104, Korea}
\affil[9]{Department of Astronomy and Space Science, Kyung Hee University, 1732 Deogyeong-daero, Giheung-gu, Yongin-si,  Gyeonggi-do 17104, Korea}
\affil[10]{McDonald Observatory, The University of Texas at Austin, 3640 Dark Sky Drive, Fort Davis, TX 79734, USA}
\affil[11]{Department of Astronomy, The University of Texas at Austin, 2515 Speedway, Austin, TX 78712, USA} 
\affil[12]{Department of Physics \& Astronomy, Rice University, 6100 Main St. MS-108, Houston, TX 77005, USA}
\affil[13]{Mitchell Institute for Fundamental Physics and Astronomy and Department of Physics and Astronomy, Texas A\&M University, College Station, TX 77843-4242, USA}
\affil[14]{Phillips Academy, 180 Main St, Andover, MA, 01810, USA}
\affil[15]{Department of Earth Science Education, Seoul National University, Gwanak-gu, Seoul 151-742, Korea}
\affil[16]{Department of Astronomy, Yonsei University, 50 Yonsei-ro, Seodaemun-gu, Seoul 03722, Korea}
\affil[17]{Department of Astronomy and Space Science, Sejong University, 209 Neungdong-ro, Kwangjin-gu, Seoul 05006, Korea}
\def\corrauthor{
M. Im
}

\def\runningauthor{
M. Im, C. Choi, S. Hwang, G. Lim, et al. 
}

\def\runningtitle{
IMSNG
}

\def\keywords{
journals: individual --- journals: JKAS
}

\def\abstracttext{
 Intensive Monitoring Survey of Nearby Galaxies (IMSNG) is a high cadence 
 observation program monitoring nearby galaxies with high probabilities of hosting supernovae (SNe). 
  IMSNG aims to constrain the SN explosion mechanism by inferring sizes of SN progenitor systems through the detection of  the shock-heated emission that lasts less than a few days after the SN explosion. To catch the signal, IMSNG utilizes a network of 0.5-m to 1-m class telescopes around the world and monitors the images of 60 nearby galaxies at distances $D < 50$ Mpc to a cadence as short as a few hours. The target galaxies are bright in near-ultraviolet (NUV) with $M_{NUV} < -18.4$ AB mag and have high probabilities of hosting SNe (0.06 SN yr$^{-1}$ per galaxy). With this strategy, we expect to detect the early light curves of 3.4 SNe per year to a depth of $R \sim 19.5$ mag, enabling us to detect the shock-heated emission from a progenitor star with a radius as small as 0.1 $R_{\odot}$. The accumulated data will be also useful for studying faint features around the target galaxies and other science projects. So far, 18 SNe have occurred in our target fields (16 in IMSNG galaxies) over 5 years, confirming our SN rate estimate of 0.06 SN yr$^{-1}$ per galaxy.
}

\begin{document}
\jkashead %% set title, authors, abstract, etc.

%%%%%%%%%%%%%%%%%%%%%%%%%%%%%%%%%%%%%%%%%%%%%%%%%%%%%%%%%%%%%%%%%%%%%
%%% BEGIN MAIN TEXT HERE %%%%%%%%%%%%%%%%%%%%%%%%%%%%%%%%%%%%%%%%%%%%
%%%%%%%%%%%%%%%%%%%%%%%%%%%%%%%%%%%%%%%%%%%%%%%%%%%%%%%%%%%%%%%%%%%%%

\section{Introduction} \label{sec:intro}

  Many stars die with dramatic explosions, which we call supernovae (SNe). Understanding the SN explosion has several implications of significant astrophysical importance. For example, type Ia SNe (SNe Ia) have been used as a key distance indicator to understand the expansion of the universe, but the theoretical reason behind the success of SNe Ia as a distance indicator is yet to be clarified. SNe also signify the end of the life of stars and detailed knowledge on SNe completes our understanding on the stellar evolution.  
  
 The SN explosion mechanism has been well formulated theoretically. Core-collapse SNe (CC SNe) are neutrino-powered explosions in massive stars and SNe Ia are thermonuclear explosions of white dwarfs (WDs) in close binary systems \citep{Branch2017}.
 Studies on light curves, spectra, neutrino emission, host galaxies, and  remnants of SNe support such an idea.
 However, direct observational evidence for the proposed SN progenitor star properties is scarce and mostly limited to bright progenitors of SNe II-P  (e.g., \citealt{Smartt2004,Fraser2011,VanDyk2012a,VanDyk2012b,VanDyk2013}), and efforts are still continuing to gather the observational proof for the SN explosion theoretical framework. Especially, what has been lacking is the evidence whether the progenitor stars have the characteristic sizes as theoretically proposed. For SNe Ia, the progenitor system is thought to be a binary star system with an exploding WD, and the companion star can be either a main sequence or a red giant star (single degenerate system; e.g., \citealt{Whelan1973,Nomoto1982,Hachisu1996,Li1997,Langer2000,Han2004}), or another WD (double degenerate system; e.g.,\citealt{Webbink1984,Iben1984,Yoon2007,Pakmor2012}). The thermonuclear explosion from the merger of a WD and an asymptotic giant branch (AGB) star can also occur (core degenerate system; e.g., \citealt{Sparks1974,Soker2015}). For CC SNe, a wide range of progenitors are expected, from red supergiants with $R \simeq$ several $100\,R_{\odot}$ \citep{Chun2018} to H/He envelope-stripped progenitors of 
  SNe Ib/Ic with $R \lesssim$ a few $R_{\odot}$ \citep{Yoon2010}.  
A binary origin of stripped-envelope SNe (i.e., SN IIb, Ib and Ic) has been suggested, and their progenitor sizes can have a significant diversity from several $100$ to $\sim 0.2 \,R_{\odot}$ \citep{Yoon2010,Eldridge2013,Yoon2015, Yoon2017,Ouchi2017}.

  Recently, it has been recognized that the light curve of SN shortly after its explosion contains valuable information about its progenitor system and can be used to set a limit on the progenitor size, R$_{*}$ (e.g.,\citealt{Kasen2010,Nakar2010,Rabinak2011,Piro2013,Piro2014,Piro2016,Noebauer2017}).  The information is contained in the shock-heated emission that appears shortly after the explosion from the outermost layers of SN ejecta, and, if the progenitor is in a binary system,  the companion star that is affected by SN shock.  
  The brightness of the shock-heated emission is proportional to the size of the exploding star (CC SNe) and/or the companion star (SNe Ia). This emission lasts for only a few hours to a few days after the explosion and it is expected to be only $R = -12$ AB mag or fainter for a double degenerate binary progenitor with sizes of 0.01 to 0.1 $R_{\odot}$ \citep{Yoon2007,Pakmor2012,Tanikawa2015}, but could be brighter for massive stars at $R = -14$ to  $-16$ AB mag for progenitor sizes of 1.0 to 10.0 $R_{\odot}$ (see Figure \ref{fig:lc_model}).
  
   Despite the difficulties associated with catching the SN shock-heated emission, several groups succeeded in detecting the emission or setting the upper limits on the progenitor stars.
  For SNe IIb, \citet{Bersten2012} analyzed the early light curve of SN 2011dh (SN IIb) and concluded that its progenitor size is $R \simeq 200 \, R_{\odot}$. The HST images taken two years after the explosion revealed that the progenitor was indeed a yellow supergiant star with an extended envelope \citep{VanDyk2013}.  
   For SNe Ia, it has been proven more challenging to catch the shock-heated emission probably due to the fact that the companion star appears to be compact.
   For example, limits on stellar sizes have been obtained for several SNe Ia from early light curves such as SN 2009ig \citep{Foley2012}, SN 2011fe \citep{Nugent2011, Bloom2012}, SN 2012cg \citep{Silverman2012,Shappee2018}, SN 2012ht \citep{Yamanaka2014}, SN 2013dy \citep{Zheng2013}, and a number of low redshift SNe Ia studied by the {\it Kepler} mission \citep{Olling2015}, all pointing toward small-sized companion stars. However, several results exist that allow a larger companion star.  \citet{Im2015b} caught the very early light curve of SN 2015F at 23.9 Mpc using the Lee Sang Gak Telescope (LSGT; Im et al. 2015a). The detection of a possible shock-heated emission suggests that the companion star size is $\lesssim 1 \,R_{\odot}$. \citet{Goobar2015} studied the shape of the early light curve of SN 2014J, suggesting a large progenitor for this SN Ia. \citet{Cao2015b} claimed a detection of UV flash in the early light curve of another SN Ia, iPTF14atg, again suggesting a very large companion star. \citet{Hosseinzadeh2017} analyzed the early light curve of SN 2017cbv showing that the best fit is obtained when assuming a companion star with a radius of $R = 56\,R_{\odot}$. However, they also caution that the blue excess emission could be due to other mechanisms such as circumstellar material interaction. For a review on this subject on both theoretical and observational sides, see \citet{Maeda2016}. Constraints on SNe Ib/Ic are much scarcer than SNe Ia, mostly due to the lack of early light curve data. 
   
  To catch the shock-heated emission in the SN early light curve, we are conducting a monitoring survey of nearby galaxies using 1-m class telescopes around the world. The sample is made of galaxies at distances  less than $50$ Mpc. The galaxies are chosen to be those that have relatively high probabilities of hosting SNe due to its high star formation rate (SFR) and low extinction. The cadence of the observation can be as short as a few hours, but nominally about 1 day. In this paper, we describe this survey, the Intensive Monitoring Survey of Nearby Galaxies (IMSNG).  Section \ref{sec:target} describes the target selection criteria, Section \ref{sec:facilities} describes the observational facilities used by the survey, and Section \ref{sec:snrate} examines the expected rate of SNe in the IMSNG galaxies. 
 The SNe and other transients that occurred in IMSNG galaxies are given in 
 Section \ref{sec:imsng-transients}, and additional science cases of IMSNG are presented in  
   Section \ref{sec:science-cases}. Finally,   Section \ref{sec:summary} gives 
   the summary and future prospects of the survey.

\section{Target Selection} \label{sec:target}

  In general, hundreds of galaxies 
 need to be monitored every night to catch the early light curves of a few SNe per
 year,  since SN rate (SN yr$^{-1}$ per galaxy ) is on average of order of 0.01 SN yr$^{-1}$ \citep[e.g., see ][]{Graur2017a}.  This, however, is not a practical approach since there is a limit on the number of galaxies that can be covered by a single telescope at a given site. 
 
 In order to increase the chance of catching SNe, we examined how host galaxy properties 
 influence the SN rate. It has been known that galaxy properties such as SFR, stellar mass, and specific SFR affect the SN rate \citep[e.g.,][]{Botticella2012,Gao2013,Graur2017a,Graur2017b}.
 CC SNe occur in massive stars that trace star formation activities \citep[e.g.,][]{Botticella2012}. Similarly, 
  recent studies suggest that SNe Ia also occur more often in galaxies with higher SFRs \citep[e.g.,][]{Smith2012,Gao2013,Botticella2017}, with the SN rate being about 10 times higher in star forming galaxies than in passively evolving galaxies \citep{Smith2012} and the SN rate increasing proportionally with SFR \citep{Smith2012, Gao2013}. This is so because the distribution of the delay time between the WD detonation and the time of the binary formation is proportional to $t^{-1}$, preferring SNe Ia with short delay times. Discussions on the delay time distribution and the dependence on SFR can be found in \citet{Maoz2012} and references therein. 

  Taking into account these factors, we selected our targets to be galaxies based on the following criteria. 
  Note that $M_{\rm NUV}$ is corrected for the Galactic extinction \citep{Bai2015}.
  
\begin{enumerate}
\item $M_{\rm NUV} < -18.4$ AB mag
\item $D < 50$ Mpc
\item $b > 20$ degree
\end{enumerate}

    The first criterion about the NUV magnitude is driven by the fact that SNe occur more frequently in galaxies with high SFR, and NUV is a good proxy for SFR. The NUV magnitude cut roughly corresponds to an SFR of 1 $M_{\odot}$ yr$^{-1}$.  
 Furthermore, the NUV selection preferentially selects high SFR with  little internal extinction. This way, we can expect to detect the shock-heated emission without worrying much about the dust extinction correction that can complicate the light curve analysis. As we shall show later, this NUV magnitude cut provides a galaxy sample whose mean SN rate is about 0.06 SN yr$^{-1}$ -- about a factor of six increase above the canonical SN rate.
   
% The second consideration is the likelihood of detecting SNe that can be studied easily 
% without getting affected by dust extinction. Although star formation rate is a key parameter
% controlling SN rate, it is no use if we cannot detect the new SN hidden in dust-enshrouded
% star forming regions. This aspect is very important when studying the shock-heated emission
% in the early light curve since such an emission is expected to be faint and can be easily hidden
% by dust.  Furthremore, heavy dust extinction complicates the light curve
% analysis due to the uncertainty related to the dust extinction correction.
 
\begin{figure}[h]
\centering
\includegraphics[width=87mm]{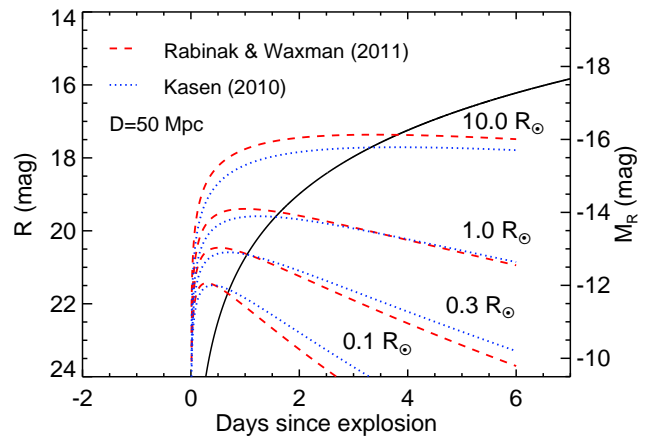}
\caption{Model predictions of the shock-heated emission light curves at 50 Mpc, 
 overlayed on the best-fit early light curve of SN 2015F \citep{Im2015b} that is fitted to the data after $\sim 1$ day after the explosion and shifted to 50 Mpc (the solid black line).  The best-fit early light curve of SN 2015F represents a typical SN Ia light curve due to radioactive decay. 
The dashed lines are for \citet{Rabinak2011} for a CC SN, and the dotted lines are for \citet{Kasen2010} due to the shock-heated emission from a companion star in SN Ia. The shock-heated emission from SN Ia is expected to be anisotropic, and can be fainter by 2.5 mag. The case plotted here is for the most optimal viewing angle. 
%The radioactive decay SN light curve follows the best-fit curve of SN 2015F \citep{Im2015b}, which is shifted to 50 Mpc. 
\label{fig:lc_model}}
\end{figure}

  The second criterion of $D < 50$ Mpc makes it possible to detect the shock-heated emission of SNe with small progenitors. Figure \ref{fig:lc_model} shows the light curve of the shock-heated emission for various sizes of progenitor stars at $D=50$ Mpc, and a typical SN Ia light curve without shock-heated emission.  Two model curve sets are plotted, one for CC SN by \citet{Rabinak2011} and another for a companion star in an SN Ia of \citet{Kasen2010}. The model parameters are adopted as in \citet{Im2015b}. Note that the shock-heated emission from SNe Ia can be anisotropic, and can be about 10 times weaker than the case for the optimal viewing angle which is plotted in Figure \ref{fig:lc_model}. 
  With the IMSNG depth of $R = 19.5$ magnitude, Figure 1 demonstrates that we can theoretically expect the detection of  the shock-heated emission from a progenitor size of  $\sim 1~R_{\odot}$ at $D = 50$ Mpc, under the most optimal condition on the viewing angle and/or the timing of the observation. At 20 Mpc, we can reach 2 magnitude fainter, and possibly detect the shock-heated emission for a progenitor with $\sim 0.1 R_{\odot}$.   Note that merger of a binary WD system may occur in a common envelope that remains during the stellar evolution, and such an event would produce shock-heated emission signal identical to an exploding $\sim 0.1 R_{\odot}$ star \citep{Pakmor2012,Tanikawa2015}.  There are uncertainties in the models too, but considering that the median distance to the IMSNG galaxies is about 30 Mpc, our goal of constraining the progenitor size down to  $\sim 1~R_{\odot}$ should be realistic for many of the IMSNG galaxies if not all.
 
  For the shock-heated emission from SNe Ia, there is a degeneracy between the viewing angle and the progenitor size. It is not clear yet how the degeneracy can be broken from the shock-heated emission alone, but we expect to be able to statistically infer the mean size of SNe Ia companion stars once we accumulate a large number of shock-heated emission data ($> 10$ objects).
   
 %%% FIGURE %%%%%%%%%%%%%%%%%%%%%%%%%%%%%%%%%%%%%%%%%%%%%%%%%%%%%%%%%%%%%%%%%%%%
%%%%%%%%%%%%%%%%%%%%%%%%%%%%%%%%%%%%%%%%%%%%%%%%%%%%%%%%%%%%%%%%%%%%%%%%%%%%%%%  

 The third criterion is imposed in order to avoid the heavy Galactic extinction and contamination from stars in our Galaxy. However, several galaxies at low Galactic latitude are found to be prolific in SNe or quite bright in $M_{\rm NUV}$. For this reason, we made two exceptions, and included NGC 6946 and ESO 182-G10 in our sample. 

Based on these criteria, we selected galaxies for our monitoring study from the Galaxy Evolution Explorer (GALEX) UV atlas of \citet{gildepaz07} and \citet{Bai2015}, where the list of \cite{Bai2015} is a more extended version of \citet{gildepaz07}. We started our selection originally from the \citet{gildepaz07} list and created a list of 46 target galaxies, but now settled on 60 galaxies after expanding the list using the \citet{Bai2015} catalog.  
 Note that there are 22 active galactic nuclei (AGNs) in the 60 IMSNG galaxies\footnote{The AGNs are identified through the NASA/IPAC Extragalactic Database at http://ned.ipac.caltech.edu.} -- two Seyfert 1.5's, nine Seyfert 2's, ten low-ionization nuclear emission line regions (LINERs), and one  obscured low luminosity AGN (oLLAGN). So, all of them have either obscured nuclei or low level of AGN, and we can consider their UV luminosities to be dominated by star formation.

\begin{figure}[h!]
\centering
\includegraphics[width=87mm]{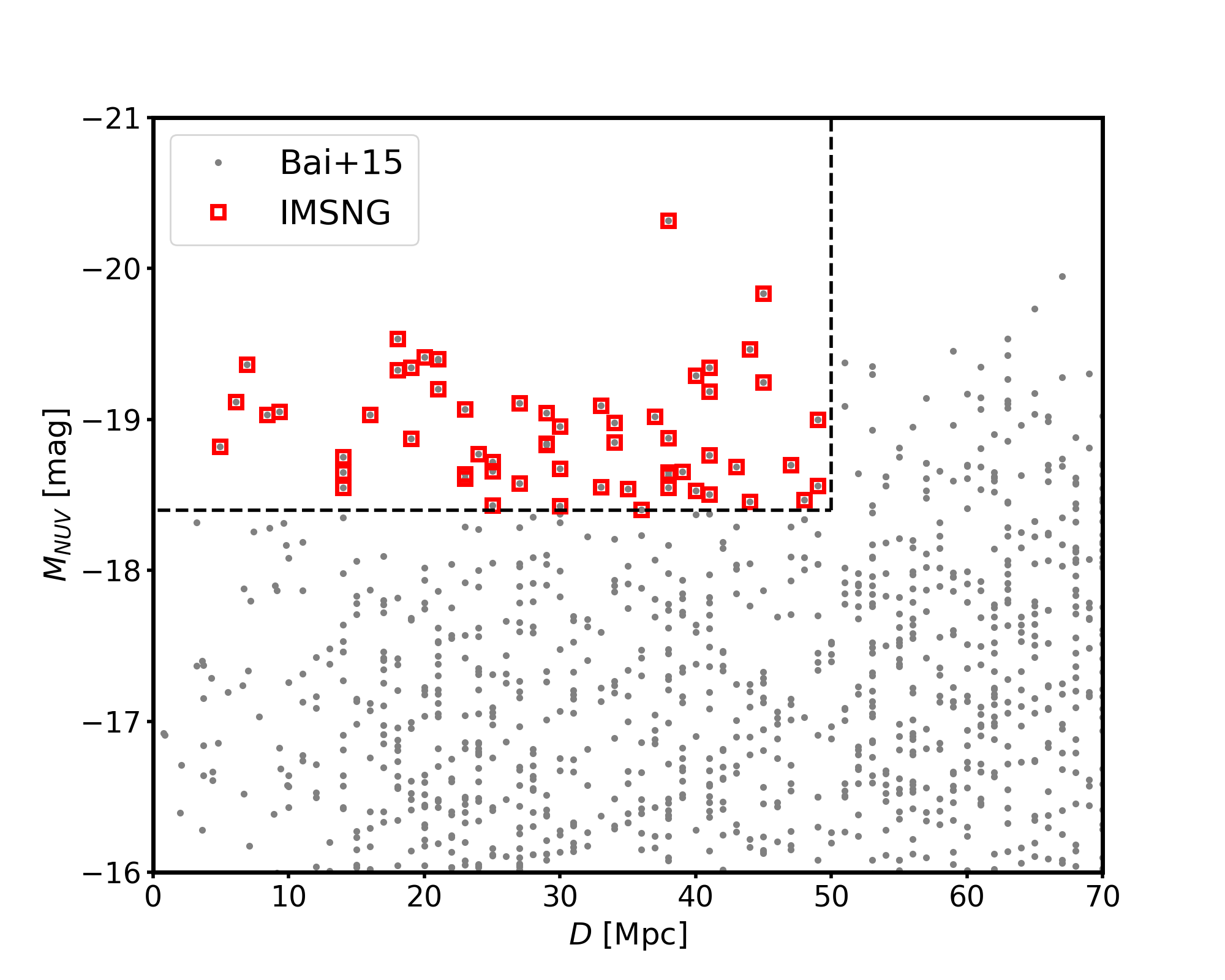}
\caption{
$M_{\rm NUV}$ (AB) versus distance (Mpc) of IMSNG galaxies (red squares), plotted over galaxies from \citet{Bai2015} (gray circles).  The area within the black dashed line denotes the region where we selected IMSNG galaxies. \label{fig:sample}}
\end{figure}

The list of the 60 IMSNG galaxies is given in Table~\ref{tab:imsng_target}. Figure \ref{fig:sample} shows the $M_{\rm NUV}$ and distance, $D$, of the IMSNG galaxies.

%%% FIGURE %%%%%%%%%%%%%%%%%%%%%%%%%%%%%%%%%%%%%%%%%%%%%%%%%%%%%%%%%%%%%%%%%%%%
%%%%%%%%%%%%%%%%%%%%%%%%%%%%%%%%%%%%%%%%%%%%%%%%%%%%%%%%%%%%%%%%%%%%%%%%%%%%%%%  

% Table: Target Table

% Galaxy name   distance    NUV mag     Morphology      Past SNe

%\clearpage

%%% TABLE %%%%%%%%%%%%%%%%%%%%%%%%%%%%%%%%%%%%%%%%%%%%%%%%%%%%%%%%%%%%%%%%%%%%%
\begin{table*}[h!]
%\begin{longtable}[t!]{| l | c | c | r | c |}
\caption{IMSNG Target Galaxies\label{tab:imsng_target}}
\centering
%\begin{multicols}{2}
%\tablehead{  & IMSNG &  Target & Galaxies  &  } \\
%\begin{minipage}[b]{0.5\textwidth}
\begin{tabular}{lccrcl}
\toprule
Name [AGN type]  & RA      & Dec      & $\rm{D_{L}}$ & NUV  & Past SNe \\ 
      & (J2000) & (J2000)  & (Mpc)        & (AB)   &    \\ 
(1)   & (2)     & (3)      & (4)          & (5)   &  (6)  \\ 
\midrule
NGC 289              &   00:52:42.348   &   $-31$:12:20.92   &   24.0   &   -18.77  &   \\ 
NGC 337  [LIN]    &   00:59:50.100   &   $-07$:34:41.45   &   23.0   &   -18.64  & 1998dn, 2011dq, 2014cx  \\ 
NGC 488                &   01:21:46.836   &   $+05$:15:24.48   &   38.0   &   -18.88  &  1976G, 2010eb \\ 
NGC 895                &   02:21:36.468   &   $-05$:31:17.00   &   37.0   &   -19.02   &  2003id \\ 
NGC 1097 [LIN]          &   02:46:19.092   &   $-30$:16:29.89   &   14.0   &   -18.55  &   1992bd, 1999eu, 2003B \\ 
NGC 1309               &   03:22:06.600   &   $-15$:24:00.07   &   29.0   &   -19.04 &  2002fk, 2012Z \\ 
NGC 1365  [S1.5]         &   03:33:36.396   &   $-36$:08:25.84   &   18.0   &   -19.33  &  1957C, 1983V, 2001du, 2012fr\\ 
UGC 2855               &   03:48:20.736   &   $+70$:07:58.30   &   14.0   &   -18.75  & 2014dg \\ 
NGC 1672  [S2]         &   04:45:42.516   &   $-59$:14:50.42   &   19.0   &   -19.34   &  2017gax \\ 
%\begin{tabular}{@{}l@{}}NGC 2207 \\ (IC 2163) \end{tabular} &   06:16:22.044   &   $-21$:22:21.76   &   38.0   &   -20.32   \\
NGC 2207/IC 2163$^{a}$          &   06:16:22.044   &   $-21$:22:21.76   &   38.0   &   -20.32 & 1975A, 1999ec, 2003H, 2010jp, 2013ai, 2018lab \\ 
%(IC 2163)               &                           &    &    &  \\
NGC 2336  [S2]        &   07:27:04.068   &   $+80$:10:41.02   &   29.0   &   -18.83  & 1987L \\ 
NGC 2442  [LIN]         &   07:36:23.796   &   $-69$:31:50.70   &   21.0   &   -19.20  & 1999ga, 2015F \\ 
NGC 2775               &   09:10:20.100   &   $+07$:02:17.23   &   43.0   &   -18.69  & 1993Z  \\ 
NGC 2776                &   09:12:14.508   &   $+44$:57:17.53   &   41.0   &   -19.34  &  \\ 
NGC 2782  [oLLAGN]       &   09:14:05.064   &   $+40$:06:49.57   &   41.0   &   -18.76  & 1994ak \\ 
NGC 2993/2992 [S2]$^{a}$   &   09:45:48.312   &   $-14$:22:06.17   &   34.0   &   -18.85  &  2003ao, AT2017ejx \\ 
IC 2537                &   10:03:51.876   &   $-27$:34:14.81   &   36.0   &   -18.40  &  2010lm \\ 
NGC 3147 [S2]            &   10:16:53.688   &   $+73$:24:02.63   &   40.0   &   -19.29  &  1972H, 1997bq, 2006gi, 2008fv \\ 
NGC 3169 [LIN]      &   10:14:14.892   &   $+03$:27:58.86   &   45.0   &   -19.25  & 1984E, 2003cg \\ 
NGC 3183               &   10:21:48.960   &   $+74$:10:37.16   &   49.0   &   -18.56  &  \\ 
NGC 3244               &   10:25:28.848   &   $-39$:49:39.00   &   38.0   &   -18.63  & 2010ev \\ 
NGC 3294               &   10:36:16.236   &   $+37$:19:28.63   &   30.0   &   -18.43  &  1990H, 1992G \\ 
NGC 3344               &   10:43:31.116   &   $+24$:55:19.74   &   20.0   &   -19.42  & 2012fh \\ 
NGC 3367  [S2]        &   10:46:35.004   &   $+13$:45:02.09   &   45.0   &   -19.84  &  1986A, 1992C, 2003aa, 2007am, 2018kp\\ 
NGC 3359               &   10:46:36.840   &   $+63$:13:26.83   &   23.0   &   -19.07  & 1985H  \\ 
NGC 3445               &   10:54:35.712   &   $+56$:59:23.32   &   33.0   &   -18.55  &  \\ 
NGC 3629               &   11:20:31.776   &   $+26$:57:47.84   &   38.0   &   -18.55  &  \\ 
NGC 3646               &   11:21:43.092   &   $+20$:10:11.10   &   44.0   &   -19.47  & 1989N, 1999cd \\ 
NGC 3938               &   11:52:49.368   &   $+44$:07:14.88   &   19.0   &   -18.87  & 1961L, 1964I, 2005ay, 2017ein \\ 
NGC 4030               &   12:00:23.580   &   $-01$:06:00.00   &   27.0   &   -19.11  &  2007aa \\ 
NGC 4038 (Arp 244)       &   12:01:53.004   &   $-18$:52:04.76   &   21.0   &   -19.40  & 1921A, 1974E, 2004gt, 2007sr , 2013dk\\ 
NGC 4039 (Arp 244)       &   12:01:53.616   &   $-18$:53:11.11   &   21.0   &   -19.39  &  \\ 
NGC 4108               &   12:06:44.316   &   $+67$:09:46.12   &   41.0   &   -18.50   &  ASASSN-15lf \\ 
NGC 4254 (M 99) [LIN]     &   12:18:49.572   &   $+14$:24:59.08   &   16.0   &   -19.03  &   1967H, 1972Q, 1986I, 2014L \\ 
%\begin{tabular}{@{}l@{}} NGC 4254 \\ (M99) \end{tabular}       &   12:18:49.572   &   $+14$:24:59.08   &   16.0   &   -19.03   \\ 
%(NGC 4254)            &                           &                                    &    &  \\
NGC 4303 (M 61)  [S2]    &   12:21:54.936   &   $+04$:28:27.05   &   18.0   &   -19.54  &  1926A, 1961I, 1964F, 1999gn, 2006ov, 2008in, 2014dt \\ 
%\begin{tabular}{@{}l@{}} NGC 4303 \\ (M61) \end{tabular}      &   12:21:54.936   &   $+04$:28:27.05   &   18.0   &   -19.54   \\ 
%(NGC 4303)            &                           &                                    &    &  \\
NGC 4314  [LIN]         &   12:22:31.980   &   $+29$:53:43.48   &   44.0   &   -18.46 &  1954A   \\ 
NGC 4321 (M 100)  [LIN]        &   12:22:54.768   &   $+15$:49:18.80   &   14.0   &   -18.65  &  2006X  \\ 
NGC 4500               &   12:31:22.152   &   $+57$:57:52.81   &   48.0   &   -18.47 &  \\ 
NGC 4653               &   12:43:50.916   &   $-00$:33:40.54   &   39.0   &   -18.66  &1999gk, 2009ik  \\ 
NGC 4814               &   12:55:21.936   &   $+58$:20:38.80   &   40.0   &   -18.53  &   \\ 
NGC5194 [S2]/5195$^{a}$ (M51)   &   13:29:52.692   &   $+47$:11:42.54   &    8.4   &   -19.03  & 1945A$^{b}$, 1994I, 2005cs, 2011dh   \\ 
%\begin{tabular}{@{}l@{}} NGC5194 (M51a)\begin{tabular}{@{}l@{}} (NGC 5194/5195)   &   13:29:52.692   &   $+47$:11:42.54   &    8.4   &   -19.03   \\ 
NGC 5236 (M83)        &   13:37:00.876   &   $-29$:51:56.02   &    4.9   &   -18.82 & 1923A, 1945B, 1950B, 1957D, 1968L, 1983N \\ 
%\begin{tabular}{@{}l@{}}NGC 5236 \\ (M83) \end{tabular}        &   13:37:00.876   &   $-29$:51:56.02   &    4.9   &   -18.82   \\ 
NGC 5371  [LIN]          &   13:55:39.936   &   $+40$:27:41.90   &   33.0   &   -19.09  & 1994Y \\ 
NGC 5430               &   14:00:45.720   &   $+59$:19:42.24   &   47.0   &   -18.70  &  PTF10acbu (PSN) \\ 
%\begin{tabular}{@{}l@{}}NGC 5427 \\ (M101) \end{tabular}       &   14:03:12.600   &   $+54$:20:56.62   &    6.9   &   -19.36   \\ 
NGC 5457 (M101)      &   14:03:12.600   &   $+54$:20:56.62   &    6.9   &   -19.36  & 1909A, 1951H, 1970G, 2011fe \\ 
NGC 5584               &   14:22:23.772   &   $-00$:23:15.32   &   25.0   &   -18.43  & 1996aq, 2007af \\ 
NGC 5668               &   14:33:24.300   &   $+04$:27:01.19   &   25.0   &   -18.72 &  1952G, 1954B, 2004G \\ 
%\begin{tabular}{@{}l@{}} NGC 5850  \\ (UGC 9718) \end{tabular} &   15:07:07.644   &   $+01$:32:40.74   &   38.0   &   -18.65   \\ 
NGC 5850 [LIN] &   15:07:07.644   &   $+01$:32:40.74   &   38.0   &   -18.65  & 1987B \\ 
NGC 5962               &   15:36:31.680   &   $+16$:36:28.15   &   30.0   &   -18.68  & 2016afa, 2017ivu  \\ 
NGC 6070               &   16:09:58.680   &   $+00$:42:34.31   &   27.0   &   -18.58  &  \\ 
NGC 6555               &   18:07:49.188   &   $+17$:36:17.53   &   35.0   &   -18.54  &  \\ 
%\begin{tabular}{@{}c@{}} ESO \\182$-$IG 010 \end{tabular}         &   18:18:30.600   &   $-54$:41:39.41   &   49.0   &   -19.00   \\ 
ESO 182-G10$^{c}$        &   18:18:30.600   &   $-54$:41:39.41   &   49.0   &   -19.00  &  2006ci \\ 
NGC 6744  [LIN]          &   19:09:45.900   &   $-63$:51:27.72   &    9.3   &   -19.05  & 2005at \\ 
NGC 6814  [S1.5]           &   19:42:40.608   &   $-10$:19:25.32   &   23.0   &   -18.61  &   \\ 
NGC 6946$^{c,d}$          &   20:34:52.572   &   $+60$:09:13.57   &    6.1   &   -19.12  & 1980K, 2002hh, 2004et, 2008S, 2017eaw \\ 
NGC 6951  [S2]        &   20:37:14.088   &   $+66$:06:20.45   &   25.0   &   -18.66  &  1999el, 2000E, 2015G \\ 
NGC 7083             &   21:35:44.592   &   $-63$:54:09.79   &   34.0   &   -18.98  & 1983Y, 2009hm \\ 
NGC 7479 [S2]           &   23:04:56.676   &   $+12$:19:22.12   &   30.0   &   -18.96  & 1990U, 2009jf  \\ 
NGC 7552               &   23:16:10.776   &   $-42$:35:03.41   &   29.0   &   -18.84  & 2017bzc \\ 
NGC 7714/7715$^{a}$   &   23:36:14.112   &   $+02$:09:18.07   &   41.0   &   -19.18  &  1999dn, 2007fo \\ 
\bottomrule
\end{tabular}

%\end{minipage}
\tabnote{(1) Galaxy name. The name in the parenthesis is another notable name of the galaxy, and the AGN types in the large parentheses are S (Seyfert), LIN (LINER), and oLLAGN; (2) \& (3): Equatorial coordinates in J2000; (4) the luminosity distance; (5) NUV absolute magnitude in AB mag; (6) the past SNe in the galaxy.
\\ $^{\rm a}$ Galaxies in pair, the primary, NUV-selected galaxy number is given first;  $^{\rm b}$ In NGC 5195; $^{\rm c}$ Low Galactic latitude target; $^{\rm d}$ This object had five additional SNe before 1980: 1917A, 1939C, 1948B, 1968D,  and 1969P.
}
\end{table*}
%\end{longtable}
%%%%%%%%%%%%%%%%%%%%%%%%%%%%%%%%%%%%%%%%%%%%%%%%%%%%%%%%%%%%%%%%%%%%%%%%%%%%%%%

\section{Facilities} \label{sec:facilities}

\begin{figure}
\centering
\includegraphics[width=85mm]{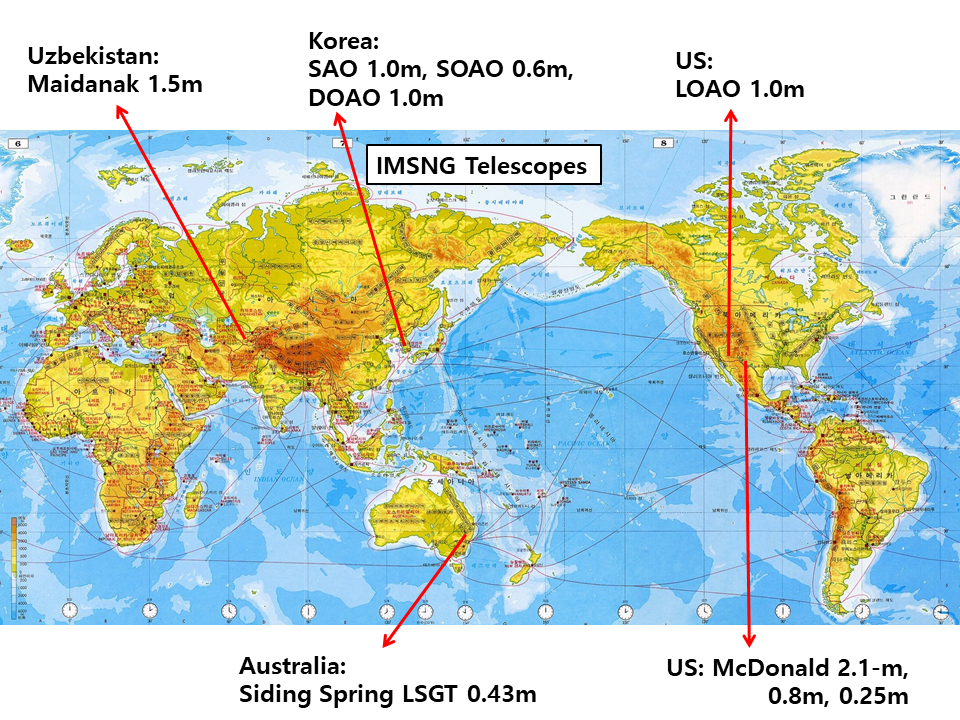}
\caption{The locations of the telescopes used by IMSNG. The background world map is taken from http://trip8.co.\label{fig:imsng_map}}
\end{figure}

 For IMSNG, we use telescopes at multiple locations around the world. This is necessary to cover different time zones and shorten the time cadence to $\sim$8 hours. 
 Table~\ref{tab:table_tel} lists our current facilities, and Figure~\ref{fig:imsng_map} shows a map where these facilities are located. Overall, most of our telescopes have field of views of order of 15 to 30 arcmin sizes, except for one wide-field 0.25m telescope with a $2.34 \degree \times 2.34 \degree$ field of view. This 0.25m telescope is a piggyback system on the McDonald Observatory's 0.8m telescope.

%%% FIGURE %%%%%%%%%%%%%%%%%%%%%%%%%%%%%%%%%%%%%%%%%%%%%%%%%%%%%%%%%%%%%%%%%%%%
%%%%%%%%%%%%%%%%%%%%%%%%%%%%%%%%%%%%%%%%%%%%%%%%%%%%%%%%%%%%%%%%%%%%%%%%%%%%%%%

%%% TABLE %%%%%%%%%%%%%%%%%%%%%%%%%%%%%%%%%%%%%%%%%%%%%%%%%%%%%%%%%%%%%%%%%%%%%
%\landscape
\begin{table*}[t]
\caption{The current list of telescopes in the IMSNG network \label{tab:table_tel}}
\centering
\begin{tabular}{lcccc}
\toprule
Observatory/ &   Instrument   &   Imager            & Longitude/  & Altitude \\
Telescope     &                      &   Field of view     &   Latitude    &  (m)  \\
\midrule
\begin{tabular}{@{}l@{}}Maidanak Observatory \\ 1.5m$^{\rm a}$\end{tabular}  &   \begin{tabular}{@{}c@{}} SNUCAM$^{\rm b}$  \\ 4k x 4k \end{tabular} &  $18\farcm3 \times 18\farcm3$ &  \begin{tabular}{@{}c@{}}66:53:47E \\ 38:40:22N\end{tabular} &  2593 \\
 \\[-0.5em]
\begin{tabular}{@{}l@{}}SNU Astronomical \\ Observatory (SAO) 1m$^{\rm c}$ \end{tabular} & \begin{tabular}{@{}c@{}} SBIG STX-16803 \\ 4k x 4k\end{tabular} &  $21\farcm2 \times 21\farcm2$  & \begin{tabular}{@{}c@{}}126:57:12E\\ 37:27:25.35N \end{tabular}  &  190 \\ 
 \\[-0.5em]
\begin{tabular}{@{}l@{}}Deokheung Optical Astronomy \\ Observatory (DOAO) 1m \end{tabular} &   \begin{tabular}{@{}c@{}} SOPHIA \\ 2k x 2k\end{tabular} &  $13\farcm2 \times 13\farcm2$ & \begin{tabular}{@{}c@{}} 127:26:48E \\ 37:31:35N\end{tabular}  & 81 \\
 \\[-0.5em]
\begin{tabular}{@{}l@{}}Sobaeksan Optical Astronomy \\ Observatory (SOAO) 0.6m\end{tabular} &  \begin{tabular}{@{}c@{}} PIXIS 2048B\\  2k x 2k\end{tabular} & $17\farcm6 \times 17\farcm6$ & \begin{tabular}{@{}c@{}} 128:27:25.3 \\ 36:56:03.9N \end{tabular}  &  1340 \\
 \\[-0.5em]
\begin{tabular}{@{}l@{}}Lee Sang Gak Telescope \\ (LSGT) 0.43m$^{\rm d}$ \end{tabular} &   \begin{tabular}{@{}c@{}}SNUCAM-II$^{\rm e}$ \\ 1k x 1k \end{tabular} &  $15\farcm7 \times 15\farcm7$ & \begin{tabular}{@{}c@{}}149:03:52E \\  31:16:24S\end{tabular} &  1122 \\
 \\[-0.5em]
\begin{tabular}{@{}l@{}}Mt. Lemmon Optical Astronomy \\ Observatory (LOAO) 1m$^{\rm f}$\end{tabular} &  \begin{tabular}{@{}c@{}} ARC CCD camera\\ 4k x 4k\end{tabular} &   $28\farcm1 \times 28\farcm1$ & \begin{tabular}{@{}c@{}} 110:47:19.3W\\  32:26:32.2N\end{tabular}  &  2776  \\
 \\[-0.5em]
 \begin{tabular}{@{}l@{}}McDonald Observatory (McD) \\ Otto-Struve 2.1m\end{tabular} &  \begin{tabular}{@{}c@{}}SQUEAN/CQUEAN$^{\rm g}$ \\ 1k x 1k \end{tabular} &   $4\farcm7 \times 4\farcm7$ & \begin{tabular}{@{}c@{}}104:01:21.4W \\ 30:40:17.4N \end{tabular} &  2076 \\
 \\[-0.5em]
\begin{tabular}{@{}l@{}}McDonald Observatory (McD)\\ 0.8m\end{tabular}  &   \begin{tabular}{@{}c@{}}CCD camera \\ 2k x 2k \end{tabular} &   $46\farcm2 \times 46\farcm$2 & \begin{tabular}{@{}c@{}}104:01:19W \\ 30:40:17N \end{tabular}  & 2057 \\
 \\[-0.5em]
\begin{tabular}{@{}l@{}}McDonald Observatory (McD)\\ 0.25m\end{tabular} &  \begin{tabular}{@{}c@{}} FLI16803 \\ 4k x 4k \end{tabular} &    $2.34\degree \times 2.34\degree$ & \begin{tabular}{@{}c@{}}104:01:19W \\ 30:40:17N \end{tabular} &  2057  \\
 \\
%\midrule
%{\sc Run A:} & & & & & & & & & \\
%mean & 0.003 & 0.003 & 0.003 & 0.005 & 0.015 & 0.036 & 0.055 & 0.061 & 0.055\\
%deviation from 0 ($\sigma_{\overline x}$) & 1.261 & 1.335 & 1.250 & 1.085 & 1.272 & 1.635 & 1.774 & 1.628 & 1.302\\ \addlinespace
%{\sc Run B:} & & & & & & & & & \\
%mean & 0.004 & 0.005 & 0.005 & 0.005 & 0.003 & 0.002 & 0.002 & 0.000 & $-$0.004 \\
%deviation from 0 ($\sigma_{\overline x}$) & 1.597 & 1.996 & 1.863 & 1.300 & 0.425 & 0.173 & 0.128 & 0.000 & 0.189 \\
\bottomrule
\end{tabular}
\tabnote{
Observatories are ordered toward E in longitude from the Prime Meridian.
\\ $^{\rm a}$ \citet{Ehgamberdiev2018} 
\\ $^{\rm b}$ \citet{Im2010}
\\ $^{\rm c}$ M. Im et al. (in preparation), http://sao.snu.ac.kr
\\ $^{\rm d}$ \citet{Im2015a}
\\ $^{\rm e}$ \citet{Choi2017}
\\ $^{\rm f}$ \citet{Han2005}
\\ $^{\rm g}$ SQUEAN\citep{Kim2016,Choi2015} is the upgraded system of CQUEAN\citep{Park2012,Lim2013,Kim2011}
}
\end{table*}
%\endlandscape
%%%%%%%%%%%%%%%%%%%%%%%%%%%%%%%%%%%%%%%%%%%%%%%%%%%%%%%%%%%%%%%%%%%%%%%%%%%%%%%

 Observations have been carried out nearly every night with LSGT 0.43m, DOAO 1.0m, and the Maidanak 1.5m telescopes \citep{Im2015a,Ehgamberdiev2018}, but in other locations, the observations have been carried out during time blocks that last about one to two weeks per month. If a galaxy is monitored at Korea, US, and Uzbekistan in one day, this would give roughly an 8 hour cadence. Some equatorial targets can be covered with the telescopes in both hemispheres, and the time cadence can be as short as 2 hours  between the observations at Korea and Australia.  The exposure times per target vary between facilities from one minute to 5 minutes, and they are set to reach about 19.5 magnitude in $R$ for point-source detection. We mostly use $R$-band filter for the monitoring observation, although a combination of $B$ and $R$ filters are used at several sites. For some telescopes, $r$-band filter is used because $R$-band is not available (LSGT, and the Otto-Struve 2.1m).  For the 0.25m telescope, we use $V$-band since IMSNG galaxies are simultaneously monitored in the other bands with the 0.8m telescope. Multiple filter observations ($BVRI$ or $griz$) are initiated once an SN is identified. 
  
%  Figure: time interval between successive observations per target
  
%   dt histogram divided by the number of days.
   
   The data taken from each telescope are downloaded to the server at Seoul National University, where they are reduced and analyzed. The data analysis, the transient detection method, and the efficiency of the observation of IMSNG will be presented elsewhere (C. Choi et al. in preparation).

%\section{Performance and Data Analysis}
    
\section{SN Rate of UV-selected Galaxies} \label{sec:snrate}

% In general, hundreds of galaxies 
% need to be monitored every night to catch the early light curves of a few SNe per
% year,  since SN rate (SN/galaxy/year) is on average of order of 0.01 SN yr$^{-1}$.  This, however, is not a practical approach since there is a limit on the number of galaxies for a telescope at a given site can cover. 
 
% In order to increase the chance of catching SNe, we examined how host galaxy properties 
% influnce the SN rate. It has been known that galaxy properties such as star formation 
% rate, stellar mass, and specific star formation rates affect the SN rate  \citep[e.g.][]{boticella12}.
% Core-collapse supernovae occur in massive stars that trace star formation activities. Similarly, type Ia supernovae also occur more often in galaxies with higher star formation. 

\begin{figure}[h]
\centering
\includegraphics[width=80mm]{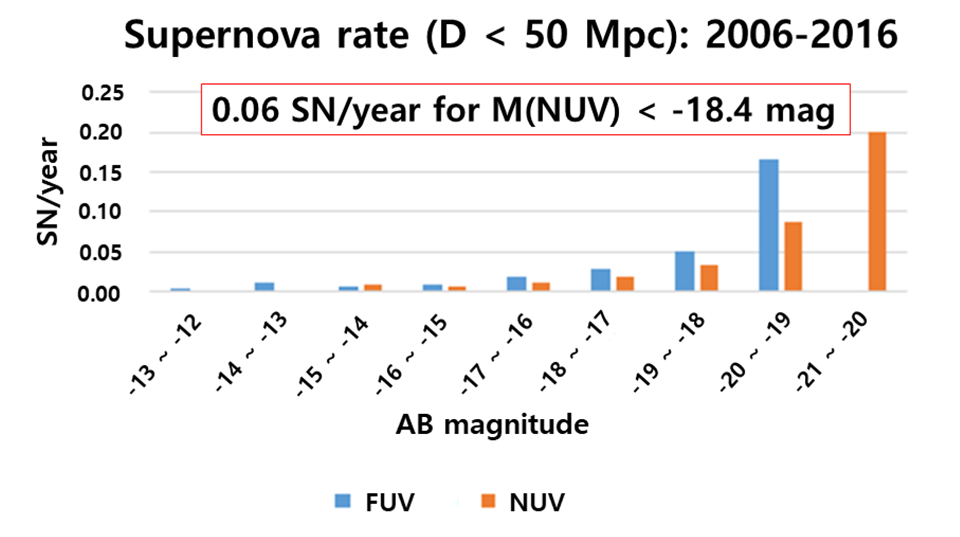}
\caption{The occurrence of SNe per year per galaxy (SN rate) as a function of FUV (blue) or NUV (red) magnitudes for galaxies within 50 Mpc. The rates were examined over the period of 2006-2016. \label{fig:SNrate}}
\end{figure}

 As we described earlier, the key to the success of our program is to select galaxies with high SN rates. We adopted the NUV selection cut for this purpose. Here, we show that SN rates are indeed high for galaxies selected this way. In Figure \ref{fig:SNrate}, we show the SN rate per year for galaxies that are selected based on GALEX NUV or far-ultraviolet (FUV) magnitudes. The UV magnitudes are taken from \citet{Bai2015}. Galaxies are limited to those at distance less than 50 Mpc for which we can detect faint shock-heated emission. Nearby galaxies have been heavily monitored over the past 10-20 years by professional and amateur astronomers, 
 and we expect that the completeness of SNe discovery rate is very high.
% COME BACK TO THIS
 %We checked the SN rate of galaxies well beyond 50 Mpc, and we confirm that the SN rate drops for the past discoveries, confirming that the completeness for the SNe discovery rate is higher at $<$ 50 Mpc.
 To estimate the SN rate, we adopted the period of 2006 to 2016 (11 years), over which the SN discovery completeness should be high thanks to various transient sky surveys. Figure \ref{fig:SNrate} shows that SN rate increases with the UV magnitude  in agreement with previous studies that SN rates are higher in galaxies with high SFR (e.g., \citealt{Botticella2012,Gao2013}). At the brightest bins in NUV, the rate can go up to 0.2 SN yr$^{-1}$. Unfortunately, there are only a few galaxies that are very bright in NUV, so we decided to bring down the magnitude limit to $M_{\rm NUV} = -18.4$ AB mag, and at such a limit, we find 0.06 SN yr$^{-1}$.     
   As we shall show in the next section, our IMSNG results in the recent 5 years agree well with the estimate from the previous 11-year data, suggesting that the completeness of the SNe sampling for the IMSNG galaxies in the 2006 to 2016 period is indeed very high. 

%%% FIGURE %%%%%%%%%%%%%%%%%%%%%%%%%%%%%%%%%%%%%%%%%%%%%%%%%%%%%%%%%%%%%%%%%%%%
%%%%%%%%%%%%%%%%%%%%%%%%%%%%%%%%%%%%%%%%%%%%%%%%%%%%%%%%%%%%%%%%%%%%%%%%%%%%%%%

%Fig: SN rate as a function of SFR + distance + Stellar mass ....
   
%IM Shall we include ``Comparison with other estimates'' ??

\section{SNe and Other Transients in IMSNG} \label{sec:imsng-transients}

As of 2018-12-31 UT, 18 SNe occurred among IMSNG galaxies or in the fields covered by IMSNG over a period of 5 years since the official start of IMSNG in 2014.  In addition, other events were also detected such as luminous red novae (LRNe) and eruptions of luminous blue variables (LBVs). These events are summarized in Table~\ref{tab:events}.  In the list, we included a special target, AT2017gfo, the optical counterpart of the gravitational wave source GW170817 \citep{Abbott2017b, Troja2017}, that was intensively observed using IMSNG facilities, although the host galaxy of AT2017gfo, NGC 4993 \citep{Im2017b}, is not in the IMSNG target list. Excluding one SN that occurred in the field of NGC 895 and another in IC 2163, 16 SNe are divided into 5 SNe Ia, 4 SNe Ib/Ic, and 7 SNe II. The current list of 16 SNe in 60 IMSNG galaxies give an SN rate of 3.2 SN yr$^{-1}$ per 60 galaxies or 0.053 yr$^{-1}$ per galaxy, which is in good agreement with the SN rate of 0.06 SN yr$^{-1}$ that is based on the 11 year period statistics. 
 The SN rates in 2016 and 2018 are low (one per year), but these low rates should be just a statistical fluctuation. The expected 1-$\sigma$ error of the yearly SN rate is $\sqrt(3.4) = 1.8$, and the low rate of one per year is only 1.3-$\sigma$ away from the mean expected rate of 3.4 SN yr$^{-1}$. 

\begin{figure}[h]
\centering
\includegraphics[width=87mm]{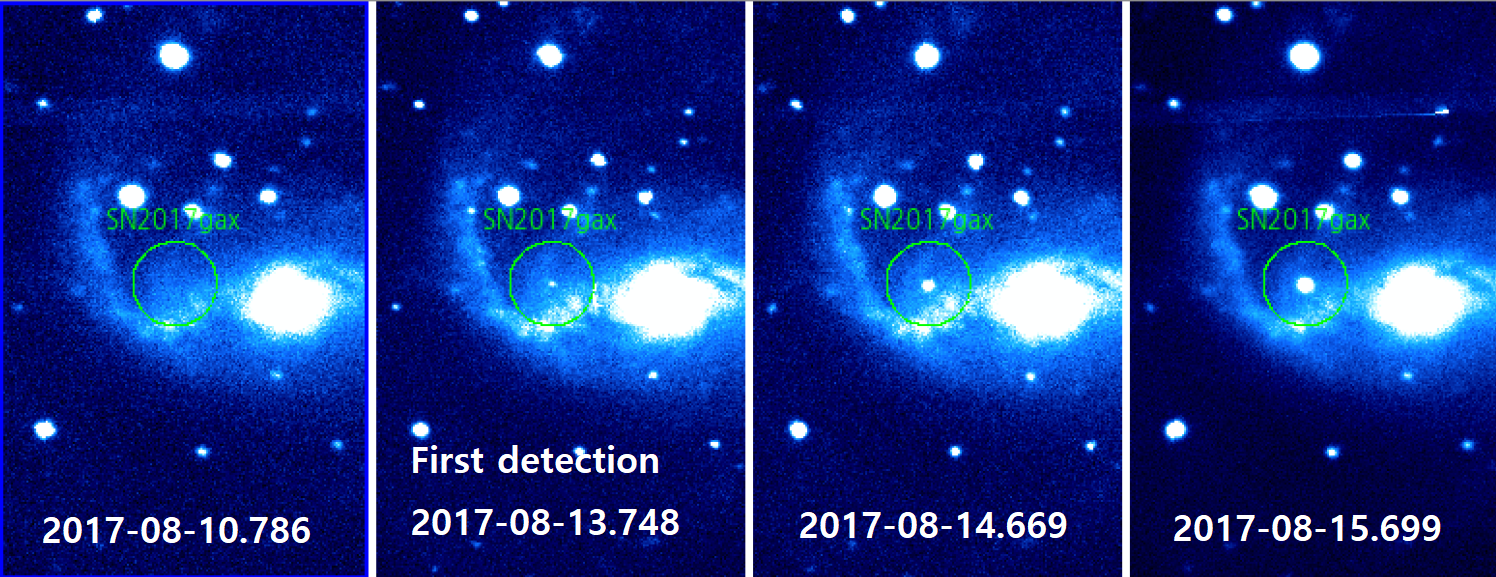}
\caption{The emergence of SN 2017gax (SN Ib/Ic) in NGC 1672 which is caught by SNUCAM-II on LSGT \citep{Im2015b,Choi2017}, one of the IMSNG telescopes \citep{Im2017a}. Each image shows a stack of three 180 sec exposure frames in $r$-band, and the green circle with a radius of $20\farcs0$ indicates the location of the SN. This example demonstrates that the high cadence IMSNG observation can catch the early optical light curves of SNe. 
 The UT date of the observation is also indicated in each image.
\label{fig:sn_example}}
\end{figure}

 Figure \ref{fig:sn_example} shows an example of an SN event that occurred in one of the IMSNG galaxies. This example shows an IMSNG image of SN 2017gax in NGC 1672 \citep{Im2017a}, before and after its explosion.  In this particular example, our first detection of the SN precedes the date of the image used for the discovery of the event by another group.

%%% FIGURE %%%%%%%%%%%%%%%%%%%%%%%%%%%%%%%%%%%%%%%%%%%%%%%%%%%%%%%%%%%%%%%%%%%%
%%%%%%%%%%%%%%%%%%%%%%%%%%%%%%%%%%%%%%%%%%%%%%%%%%%%%%%%%%%%%%%%%%%%%%%%%%%%%%%

%%% TABLE %%%%%%%%%%%%%%%%%%%%%%%%%%%%%%%%%%%%%%%%%%%%%%%%%%%%%%%%%%%%%%%%%%%%%
\begin{table*}[t]
\caption{SNe and other transients in IMSNG galaxies (2014-2018)\label{tab:events}}
\centering
\begin{tabular}{lrrl}
\toprule
Name & Type & Host & Notes \\
\midrule
2014  &     &      &    \\ 
\midrule
SN 2014L   &  Ic   &  M99/NGC 4254  &  01/26.83, gap in time cov., test obs. period\\
{\bf SN 2014cx}  &  IIP  &  NGC 337    &   09/2.275, Maidanak(08/31, 09/03)  \\  
ASASSN-14ha  & II  & NGC 1566  &  09/10.290, Before LSGT op. \\
AT2014ej & LBV & NGC 7552 & 09/24.460  \\ 
SN 2014dg & Ia  & UGC 2855 &  09/11.725, Before the addition of UGC 2855 in the list\\
SN 2014dt  &  Ia   &   M61/NGC 4303  & 10/29.838, LOAO/Maidanak out of order \\
\midrule
2015  &   &    &  \\
\midrule
PSN J14021678+5426205  &  LRN   &  M101   &  2015/01.20.803$^{a}$  \\
{\bf SN 2015F}  &  Ia  &  NGC 2442  &  03/09, LSGT daily coverage, \citet{Im2015b}  \\
SN 2015G  &  Ibn  & NGC 6951  &  03/23.788, altitude too low for LOAO when discovered  \\
ASASSN-15lf  & IIn   & NGC 4108   & 06/15.34, Maidanak/LOAO/McD \\
DES15X2kvt  & Ia?  & NGC 895 & 10/02, anonymous Ia in the field? \\
%SPIRITS15aac  &  ?   &  M83   & 10/29  \\
%SPIRITS15afp   &  ?   &  NGC 6946   &  12/03  \\
\midrule
2016  &   &   &   \\
\midrule
SNhunt306 & LBV?  &  NGC 772 &    02/10.0  \\ 
AT2016blu/Gaia16ada  &  LBV    &  NGC 4559   &  02/25.723  \\
{\bf SN 2016afa}  &  IIP   &   NGC 5962   & 02/12.958, Maidanak/LOAO/McD \\
AT2016jbu  & LBV   &  NGC 2442  & 12/01.596, LSGT  \\
\midrule
2017  &   &   &   \\
\midrule
SN 2017bzc  &  Ia   &  NGC 7552  &  03/07.219, before sample revision, some LSGT \\
%SPIRITS17fm  & ?  &  M101  &  03/08   \\
{\bf SN 2017eaw}  &  IIP  & NGC 6946  &  05/14.238, LOAO(5/14), McD(05/05), Maidanak(05/16) \\
%SPIRITS17ki,ka  &  ?  &  M83  & 05/19  \\
{\bf SN 2017ein}  &  Ic  &  NGC 3938  & 05/25.969, Maidanak(5/23, 25), DOAO(05/24), McD/LOAO \\
%SPIRITS17lb  &  ?  &  IC2163/NGC2207  &  05/28, LSGT, Maidanak  \\
{\bf SN 2017ejx} &  IIP  &  NGC 2993  & 05/30.040, LSGT(05/24,29), LOAO(05/15), DOAO(05/17)  \\
{\bf SN 2017gax} &  Ib/Ic   &  NGC 1672  &  08/14.712, LSGT(from 08/13, 14 and on) \\
AT2017gfo/GW170817  & Kilonova & NGC 4993  & 08/17.98, LSGT, KMTNet \\
SN 2017ivu  & IIP  & NGC 5962  & 12/11.857, altitude too low when discovered \\
\midrule
2018  &   &   &  \\
\midrule
{\bf SN 2018kp}  &  Ia   &  NGC 3367  & 01/24.244, daily coverage from McD, DOAO, LOAO \\
AT2018ikn & AGN flare?  &  NGC 2992  & LSGT \\
SN 2018lab & II & IC2163 & 12/29.13, bad weather \\ 
\bottomrule
\end{tabular}
\tabnote{
The information in Notes include the discovery month and dates in UT, and the primary IMSNG facilities that followed up the event. The dates of observations around the discovery dates are also indicated in the parentheses for events of interest. The object names in bold are the events of interest for which we have an extended dataset before and just after the explosion. 
\\  $^{a}$ After the discovery, the object was found to have $R=16.36$ mag at 2014-11-10 (UT), which is comparable to the brightness at  the discovery time ($V=17.5$ mag), so the burst could have occurred earlier \citep{Cao2015a}. Also, long-term light monitoring data prior to 2015 suggest that the object was a slow rising source \citep{Goranskij2016, Blagorodnova2017}. 
}
\end{table*}
%%%%%%%%%%%%%%%%%%%%%%%%%%%%%%%%%%%%%%%%%%%%%%%%%%%%

\section{Additional Science Cases} \label{sec:science-cases}

The data gathered from IMSNG are also useful for other projects too. We list several such projects here.

{\bf (1) Other stellar transients:} 
 The transients that can be discovered by IMSNG are not limited to SNe. 
 Two prominent examples are LRNe and LBV outbursts. 
 LRNe are a recently recognized class of transients with their peak luminosities somewhere between novae and SNe at $-14 < M_{V} < -6$ mag \citep{Kasliwal2012}. LRNe are suggested to be due to the merging of two stars \citep[e.g.,][]{Kulkarni2007}, although some think that these are unusually weak SNe IIP \citep{Pastorello2007}. Other physical origins for LRNe have been discussed, such as electron-capture SNe in extreme AGB stars, LBV eruptions, an asteroids crashing to WDs, accretion-induced collapses, and peculiar classical novae \citep[See references in ][]{Kasliwal2012}. 
  LBVs are another rare kind of transients arising from evolved massive stars.  LBVs undergo giant eruptions, becoming up to a few magnitudes brighter, and when they do so, they are sometimes mistakenly identified as SNe. Hence, they get the name of SN impostor.
 %  Some LBVs are known to evolve into SNe. 
 Both LRNe and LBV outbursts represent unique passages in the stellar evolution. The IMSNG data provide  a long-term light curve of the pre-brightening period as well as high cadence light curves after the brightening, which can help us learn about their progenitors and physical origin of the explosive events \citep[e.g.,][]{Blagorodnova2017}.

\begin{figure}[h]
\centering
\includegraphics[width=87mm]{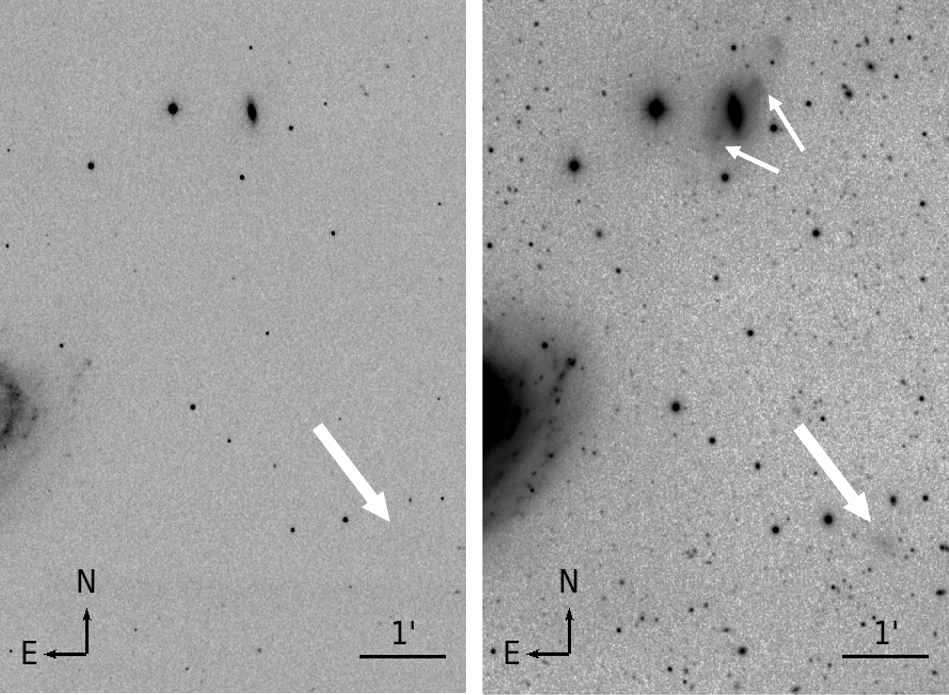}
\caption{
(Left) A single exposure (60 seconds) $R$-band image near NGC 895 galaxy. A part of NGC 895 is visible on the left. (Right) A stacked $R$-band image (2.37 hours) of the same field. The data taken from 2013 to 2016 were used. A low SB satellite galaxy candidate is marked as a large, thick arrow. Merging features are visible in the deep image for a galaxy on the top and noted with small arrows. \label{fig:satellites}
}
\end{figure}

{\bf (2) Merging features and satellite galaxies:} 
 The daily cadence data can be stacked to create deep images of nearby galaxies. The surface brightness (SB) limit of the stacked images can reach $\sim$26.5 R mag arcsec$^{-2}$ or fainter. At such SB limits, one can identify faint merging features \citep[e.g., see][]{Hong2015}, or new low SB satellite galaxies (e.g., \citealt{Park2017}). These features will be useful to understand the merging history of IMSNG galaxies, and the discovery of new satellite galaxies can help understand the problem related to the paucity of dwarf satellite galaxies with respect to the $\Lambda$CDM cosmological models.
 Figure \ref{fig:satellites} shows an example of a stacked image of 142 frames of one minute images of NGC 895 taken at the Maidanak observatory, corresponding to a total integration time of 2.37 hours, where a faint low SB satellite galaxy candidate and a merging feature around another galaxy are found in the field of NGC 895.

 %%% FIGURE %%%%%%%%%%%%%%%%%%%%%%%%%%%%%%%%%%%%%%%%%%%%%%%%%%%%%%%%%%%%%%%%%%%%
%%%%%%%%%%%%%%%%%%%%%%%%%%%%%%%%%%%%%%%%%%%%%%%%%%%%%%%%%%%%%%%%%%%%%%%%%%%%%%%  
 
 {\bf (3) Gravitational wave source optical counterpart:} 
 High cadence monitoring observation offers a unique way to uncover optical counterparts of transients such as gravitational wave (GW) sources. Localization areas of GW sources span tens to hundreds of square degrees, and it is quite common to find several IMSNG galaxies in the localization area. Optical counterparts of GW sources can be searched for IMSNG galaxies and the area surrounding them. One example is SN 2017gax in NGC 1672 which was in the localization area near GW170814,  the first GW event with the three detector operation \citep{Abbott2017a}. Eight galaxies were monitored in the localization area at the time of the GW170814 event, and one young SN, SN 2017gax was found in NGC 1672 \citep{Im2017a}. However, our monitoring observation reveals that the SN was present in the image one day before the GW event, ruling out the possibility that SN 2017gax was the GW optical counterpart (Figure \ref{fig:sn_example}). 
 
{\bf (4) AGNs:} 
 Twenty-two of our targets are known AGNs (Seyfert 1.5, Seyfert 2, and LINER galaxies, see Table \ref{tab:imsng_target}). Their nuclear activities can be traced with our survey at the IMSNG cadence. The monitoring of the nuclei of IMSNG galaxies could also reveal unexpected AGN activities that can be found only through high cadence monitoring \citep[e.g.,][]{Kim2018}. One such example is AT2018ikn which       is suggested to be a transient in a Seyfert 2 galaxy, NGC 2992, a possible AGN flare event \citep{Berton2018}.

{\bf (5) Variable stars:} High cadence observations offer an excellent opportunity to discover new variable stars in the vicinity of IMSNG galaxies. We have searched for new variable stars, and found more than a dozen cases so far \citep{Choi2018}.

{\bf (6) Asteroids:} Asteroids often appear in the IMSNG data. Their locations can be traced to provide constraints on the orbits of asteroids.

\section{Summary \& Prospects} \label{sec:summary}

 In this paper, we gave an overview of IMSNG, which is a monitoring observation project of 60 nearby galaxies at $< 50$ Mpc with a goal cadence of 8 hours. The main scientific objective of the project is to catch the early light curve of SNe, within one day from the explosion, and constraining the SN progenitor system properties such as the size of the progenitor star. Several such early light curves have been detected among IMSNG galaxies so far \citep{Im2015b}. 
 Nine 1-m class telescopes are currently being used for the monitoring observation. IMSNG galaxies are selected to be galaxies that are bright in NUV ($M_{\rm NUV} < -18.4$ mag), and we find that the SN rate of the IMSNG galaxies to be about 0.06 SN yr$^{-1}$ using SNe that appeared during the period of 2006-2016, which is six times higher than the SN rate without the UV selection. IMSNG started in 2014, and since then, there have been 16 SNe in the 60 IMSNG galaxies, confirming our SN rate estimate of 0.06 SN yr$^{-1}$.   
 With the advance of high cadence transient surveys, we expect many SNe to be discovered in their early stages. Yet, a cadence of a few hours is difficult to achieve without employing telescopes at multiple longitudes. In this regard, intensive monitoring observations with small telescopes around the world like IMSNG can cover the niche science where cadence of less than a few hours is desired to the depths of $\sim 19.5$ mag.

%%% ACKNOWLEDGMENTS (IF ANY) %%%%%%%%%%%%%%%%%%%%%%%%%%%%%%%%%%%%%%%%

\acknowledgments

We thank anonymous referees for their useful and constructive suggestions to improve the paper. This research was supported by the Basic Science Research Program through the National Research Foundation of Korea (NRF) funded by the Ministry of Education (NRF-2017R1A6A3A04005158). We thank the staffs at iTelescope.Net, DOAO, LOAO, SOAO, Maidanak \& McDonald observatories for their help with the observations and maintenance of the facilities. The research made use of the data taken with LOAO and SOAO operated by the Korea Astronomy and Space Science Institute (KASI), DOAO of National Youth Space Center (NYSC), McDonald Observatory,  Maidanak Observatory, and the Siding Spring Observatory.

%%% APPENDICES (IF ANY) %%%%%%%%%%%%%%%%%%%%%%%%%%%%%%%%%%%%%%%%%%%%%

%\appendix

%%% CALL LIST OF REFERENCES (natbib STYLE) %%%%%%%%%%%%%%%%%%%%%%%%%%

\end{document}